# Ethics of Artificial Intelligence<sup>.</sup>

---

Artificial intelligence (AI) is a digital technology that will be of major importance for the development of humanity in the near future. AI has raised fundamental questions about what we should do with such systems, what the systems themselves should do, what risks they involve and how we can control these.

After the background to the field (1), this article introduces the main debates (2), first on ethical issues that arise with AI systems as *objects*, i.e. tools made and used by humans; here, the main sections are privacy (2.1), manipulation (2.2), opacity (2.3), bias (2.4), autonomy & responsibility (2.6) and the singularity (2.7). Then we look at AI systems as *subjects*, i.e. when ethics is for the AI systems themselves in machine ethics (2.8.) and artificial moral agency (2.9). Finally we look at future developments and the concept of AI (3). For each section within these themes, we provide a general explanation of the *ethical issues*, we outline existing *positions* and *arguments*, then we analyse how this plays out with current *technologies* and finally what *policy* consequences may be drawn.

## 1    Historical and Intellectual Background

Some technologies, like nuclear power, cars or plastics, have caused ethical and political discussion and significant policy efforts to control the trajectory these technologies – usually only once some damage is done. In addition to such 'ethical concerns', new technologies challenge current norms and conceptual systems, which is of particular interest to philosophy. Finally, once we have understood a technology in its context, we need to shape our societal response, including regulation and law. All these features also exist in the case of the new technology of AI – plus the more fundamental fear that they may end the era of human control on planet Earth. The task of an article such as this is to analyse the issues, and to deflate the non-issues.

The ethics of AI and robotics has seen significant press coverage in recent years, which supports this kind of work, but also may end up undermining it: the press often talks as if the issues under discussion were just those that future technology will bring, and as though we already know what would be most ethical and how to achieve that. Press coverage thus focuses on risk, security (Brundage et al. 2018), and prediction of impact (e.g. on the job market). The result is a discussion of essentially technical problems that focus on how to achieve a desired outcome. Another result can be seen in the current discussion in policy and industry focuses on image and public relations – where the label "ethical" is really not much more than the new "green", perhaps used for "ethics washing". For a problem to qualify as a problem for AI ethics would require that we do *not* readily know what is the right thing to do. In this sense, job-loss, theft or killing with AI are not a problem in ethics, but whether these are permissible under certain circumstances *is* such a problem. This article focuses on the genuine problems of ethics where we do not readily know what the answers are.

A last caveat is in order for our presentation: The ethics of AI and robotics is a very young field within applied ethics, with significant dynamics, but few well-established issues and no authoritative overviews – though there are beginnings (Bryson 2019; European Group on Ethics in Science and New Technologies 2018; Floridi et al. 2018; Gibert 2019), and policy recommendations (AI HLEG 2019). So this article cannot just reproduce what the community has achieved thus far, but must propose an ordering where little order exists.

The notion of 'artificial intelligence' (AI) is understood broadly here, as any kind of artificial computational system that shows intelligent behaviour, i.e. complex behaviour that is conducive to reaching goals. This means we incorporate a range of machines, including those in 'technical AI' that show only limited abilities in learning or reasoning but excel at the automation of particular tasks, as well as machines in work on 'general AI' that aims at creating a generally intelligent agent.

AI somehow gets closer to our skin than other technologies – thus the field of 'philosophy of AI'. Perhaps this is because the project of AI is to create machines that have a feature central to how we humans see ourselves, namely as feeling, thinking, intelligent beings. The main purposes of an artificial intelligent agent probably involve sensing, modelling, planning and action, but current AI applications also include perception, text analysis, natural language processing (NLP), logical reasoning, game-playing, decision support systems, data analytics, predictive analytics, as well as autonomous vehicles and other forms of robotics. AI may involve any number of computational techniques to achieve these aims; be that classical symbol-manipulating AI, inspired by natural cognition, or machine learning via neural networks.

Historically, it is remarkable that the term "AI" was used as above ca. 1950-1975, then it came into disrepute during the 'AI winter', ca. 1975-1995, and narrowed. As a result, areas such as 'machine learning', 'natural language processing' and 'data science' were often not labelled as 'AI'. Since, ca. 2010, the use has broadened again, and at times almost all of computer science and even high-tech is lumped under 'AI'. Now a name to be proud of, a booming industry with massive capital investment, and on the edge of hype again.

While AI can be entirely software, robots are physical machines that are subject to physical impact, typically through 'sensors', and they exert physical force onto the world, typically through 'actuators', like a gripper or a turning wheel. Accordingly, autonomous cars or planes are robots, and only a minuscule portion of robots is 'humanoid' (human-shaped), like in the movies. Some robots use AI, and some do not: Typical industrial robots blindly follow completely defined scripts with minimal sensory input and no learning or reasoning.

Policy is only one of the concerns of this article. There is significant public discussion about AI ethics, and there are frequent pronouncements from politicians that the matter requires new policy – which is easier said than done: Actual technology policy is difficult to plan and to enforce. It can take many forms, from incentives and funding, infrastructure, taxation, or good-will statements, to regulation by various actors, and the law. Policy for AI will possibly come into conflict with other aims of technology policy or general policy. Governments, parliaments, associations and industry circles in industrialised countries have produced reports and white papers in recent years, and some have generated good-will slogans ('trusted/responsible/humane/human-centred/good/beneficial AI'), but is that what is needed? For a survey, see (Jobin, Ienca, & Vayena 2019) and our list on PT-AI Policy Documents and Institutions.

For people who work in ethics and policy, there is probably a tendency to overestimate the impact and the threats from a new technology, and to underestimate how far current regulation can reach (e.g. for product liability). On the other hand, there is a tendency for businesses, the military and some public administrations to 'just talk' and do some 'ethics washing' in order to preserve a good public image and continue as before. Actually implementing legally binding regulation would challenge existing business models and practices. Actual policy is not just an implementation of ethical theory, but subject to societal power structures – and the agents that do have the power will push against anything that restricts them. There is thus a significant risk that regulation will remain toothless in the face of economical and political power.

Though very little actual policy has been produced, there are some notable beginnings: The latest EU policy document suggests 'trustworthy AI' should be lawful, ethical and technically robust, and then spells this out as seven requirements: human oversight, technical robustness, privacy and data governance, transparency, fairness, well-being and accountability (AI HLEG 2019). Much European research now runs under the slogan of 'responsible research and innovation' (RRI) and 'technology assessment' has been a standard field since the advent of nuclear power. Professional ethics is also a standard field in information technology, and this includes issues that are relevant here. We also expect that much policy will eventually cover specific uses or technologies of AI and robotics, rather than the field as a whole (Calo 2018). In this article, we discuss the policy for each type of issue separately, rather than for AI or robotics in general.

The issue of which impact AI will have on employment is not discussed in this article since there is a separate entry in this Handbook on the topic.

## 2    Main Debates

In this section we outline the ethical issues of human use of AI and robotics systems that can be more or less autonomous – which means we look at issues that arise with certain uses of the technologies, which would not arise with others. It must be kept in mind, however, that technologies will always cause some uses to be easier and thus more frequent, and hinder other uses. The design of technical artefacts thus has ethical relevance for their us, so beyond 'responsible use', we also need 'responsible design' in this field. The focus on use does not presuppose which ethical approaches are best suited for tackling these issues; they might well be virtue ethics rather than consequentialist or value-based. This section is also neutral with respect to the question whether AI systems truly have 'intelligence' or other mental properties: It would apply equally well if AI and robotics are merely seen as the current face of automation.

### 2.1    Privacy

There is a general discussion about privacy and surveillance in information technology (e.g. Macnish 2017; Roessler 2017), which mainly concerns the access to private data and data that is personally identifiable. Privacy has several well-recognised aspects, e.g. 'the right to be let alone', information privacy, privacy as an aspect of personhood, control over information about oneself, and the right to secrecy (Bennett & Raab 2006). Privacy studies have historically focused on state surveillance by secret services but now include surveillance by other state agents, businesses and even individuals. The technology has changed significantly in the last decades while regulation has been slow to respond (though there is the (GDPR 2016)) – the result is a certain anarchy that is exploited by the most powerful players, sometimes in plain sight, sometimes in hiding.

The digital sphere has widened greatly: All data collection and storage is now digital, our lives are increasingly digital, most digital data is connected to a single Internet, and there is more and more sensor technology in use that generates data about non-digital aspects of our lives. AI increases both the possibilities of intelligent data collection and the possibilities for data analysis. This applies to blanket surveillance of whole populations as well as to classic targeted surveillance. In addition, much of the data is traded between agents, usually for a fee. At the same time, controlling who collects which data, and who has access, is much harder in the digital world than it was in the analogue world of paper and telephone calls. Data collection, sale and use are shrouded in secrecy.

The data trail we leave behind is how our 'free' services are paid for – but we are not told about that data collection and the value of this new raw material, and we are manipulated into leaving ever more such data. The main data-collection for the 'big 5' companies (Amazon, Google/Alphabet, Microsoft, Apple, Facebook) appears to be based on deception, exploiting human weaknesses, furthering procrastination, generating addiction, and manipulation (Harris 2016). The primary focus of social media, gaming, and most of the Internet in this 'surveillance economy' is to gain, maintain and direct attention – and thus data supply.

Many new AI technologies amplify these issues. For example, face recognition in photos and videos allows identification and thus profiling and searching for individuals. This continues using other techniques for identification, e.g. 'device fingerprinting', which are commonplace on the Internet (sometimes revealed in the 'privacy policy'). Together with the 'Internet of things', the so-called 'smart' systems (phone, TV, oven, lamp, virtual assistant, home, …), the 'smart city' (Sennett 2018) and 'smart governance', robots are set to become part of the data-gathering machinery that offers more detailed data, of different types, in real time, with ever more information. These systems will often reveal facts about us that we ourselves wish to suppress or are not aware of: they know more about us than we know ourselves. With the last sentence of his bestselling book *Homo Deus* (Harari 2016) asks about the long-term consequences of AI: "What will happen to society, politics and daily life when non-conscious but highly intelligent algorithms know us better than we know ourselves?"

In effect, this surveillance and attention economy "is the business model of the Internet" (Schneier 2015), sometimes called 'surveillance capitalism' (Zuboff 2019). It has caused many attempts to escape from the grasp of these corporations, e.g. in exercises of 'minimalism', or through the open source movement, but it appears that present-day citizens have lost the degree of autonomy needed to escape while fully continuing with their life and work. We have lost ownership of our data, if 'ownership' is the right relation here.

Privacy-preserving techniques that can largely conceal the identity of persons or groups are now a standard staple in data science; they include (relative) anonymisation, access control (plus encryption) and other models where computation is carried out with fully or partially encrypted input data; in the case of 'differential privacy' this is done by adding calibrated noise to encrypt the output of queries (Abowd 2017). While requiring more effort and cost, such techniques can avoid many of the privacy issues. Some companies have also seen better privacy as a competitive advantage that can be leveraged and sold at a price.

One of the major practical difficulties is to actually enforce regulation, both on the level of the state and on the level of the individual who has a claim. They must identify the responsible legal entity, prove the action, perhaps prove intent, find a court that declares itself competent … and eventually get the court to actually enforce its decision. Well-established legal protection of rights such as consumer rights, product liability and other civil liability or protection of intellectual property rights is often missing in digital products, or hard to enforce. This means that companies with a 'digital' background are used to testing their products on the consumers, without fear of liability, while heavily defending their intellectual property rights.

In sum, we have an ever-growing data collection about users and populations – to such an extent that these systems and their owners know more about us than we know ourselves. We users are manipulated into providing data, unable to escape this data collection and without knowledge of data access and use. We are not even able to enforce our legal rights because are unable to identify a legal entity and hold it accountable. We have lost control. The surveillance economy is a scandal that still has not received due public attention

### 2.2 Manipulation

The ethical issues of AI in surveillance go beyond the mere *accumulation* of data and direction of attention: They include the *use* of information for problematic purposes. On of these is the manipulation behaviour, online and offline – mostly aiming at user's money. Of course, efforts to manipulate behaviour in a way that undermines autonomous rational choice are ancient, but they may gain a new quality when they use AI systems. Given users' intense interaction with data systems and the deep knowledge about individuals this provides, we are vulnerable to 'nudges', manipulation and deception. With sufficient prior data, algorithms can be used to target individuals or small groups with just the kind of input that is likely to influence these particular individuals.

Many advertisers, marketers and online sellers will use any legal means at their disposal, including exploitation of behavioural biases, deception, and the generation of addiction (Costa & Halpern 2019) – e.g. through 'dark patterns' on web pages or in games (Mathur et al. 2019). Such manipulation is the business model in much of the gambling and gaming industries, but it is spreading, e.g. to low-cost airlines. Gambling and the sale of addictive substances are highly regulated, but online manipulation and addiction is not. Manipulation of online behaviour is becoming a core business model of the Internet.

Furthermore, social media are now the prime locations for political propaganda. This influence can be used to steer voting behaviour, as in the Facebook-Cambridge Analytica 'scandal' (Woolley & Howard 2017) and – if successful – it may harm the autonomy of individuals (Susser, Roessler, & Nissenbaum 2019).

Improved AI 'faking' technologies make what once was reliable evidence into unreliable evidence – this has already happened to digital photos, sound recordings and video … and it will soon be quite easy to create (rather than alter) 'deep fake' text, photos and video material with any content desired. Soon, sophisticated real-time interaction with persons over texting, phone or video will be faked, too. So we cannot trust digital interaction, while we are at the same time increasingly dependent on such interaction.

The policy in this field of privacy and manipulation is struggling to catch up with technical and social developments. Civil liberties and the protection of individual rights are under intense pressure from business' lobbying, secret services and other state agencies that depend on surveillance. Actual legal protection from surveillance and manipulation has diminished massively as compared to the pre-digital age (of letters, analogue telephone and oral conversation). While the EU General Data Protection Regulation (GDPR 2016) has strengthened privacy protection somewhat, the US and China prefer growth with less regulation, likely in the hope that this provides a competitive advantage. It is clear that state and business actors have increased their ability to watch and to manipulate people with the help of AI technology and will continue to do so to further their particular interests – unless reined in by policy in the interest of general society.

### 2.3 Opacity

Opacity and bias are central issues in what is now sometimes called 'data ethics' or 'big data ethics' (Floridi & Taddeo 2016). Automated AI decision support systems and 'predictive analytics' operate on data and produce a decision as 'output'. This output may range from the relatively trivial to the highly significant: "this restaurant matches your preferences", "the patient in this X-ray has completed bone growth", "application to credit card declined", "donor organ will be given to another patient", "bail is denied", or "target identified and engaged". Data analysis is often used in 'predictive analytics' in business, healthcare and other fields, to foresee future developments – since prediction is easier with AI, it will also become a cheaper commodity.

It appears that AI systems for automated decision support are part of a power structure where it will often be impossible for the affected person to know how the system came to this output, i.e. the system is 'opaque' to that person. If the system involves machine learning, it will typically be opaque even to the expert, who will not know how a particular pattern was identified, or even what the pattern is. Bias in decision systems and data sets is exacerbated by this opacity. So, at least in the cases where there is a desire to remove bias, the analysis of opacity and bias go hand in hand, and the political response has to tackle both issues together.

Many AI systems rely on machine learning techniques in (simulated) neural networks that will extract patterns from a given dataset, with or without 'correct' solutions provided; i.e. supervised, semi-supervised or unsupervised. With these techniques, the 'learning' captures patterns in the data and these are labelled in a way that appears useful to the decision the system makes, while the programmer does not really know *which* patterns in the data the system has used. In fact the programs are evolving, so when new data comes in, or new feedback is given ("this was correct", "this was incorrect"), the patterns used by the learning system change. What this means is that the outcome is not transparent to the user or programmers: It is opaque.

There are several technical activities that aim at 'explainable AI' and, more recently, a DARPA programme (Gunning 2018) and the AI4EU project on 'human-centred AI' (AI4EU 2019, 100-187, ). This does not mean that we expect an AI to 'explain its reasoning' – doing so would require far more serious moral autonomy than we currently attribute to AI systems (see below 3.2). In the EU, some of these issues have been taken into account with the (GDPR 2016), which foresees that consumers, when faced with a decision based on data processing, will have a legal "right to explanation" – how far this goes and to what extent it can be enforced is disputed (Wachter, Mittelstadt, & Russell 2018). (Zerilli, Knott, Maclaurin, & Gavaghan 2019) argue that there may be a double standard here, where we demand a high level of explanation for machine-based decisions despite the abilities of humans to explain and provide reasons sometimes not reaching that standard themselves.

The politician Henry Kissinger pointed out that there is a fundamental problem for democratic decision-making if we rely on a system that is supposedly superior to humans, but cannot explain its decisions. He says we may have "generated a potentially dominating technology in search of a guiding philosophy" (Kissinger 2018). The political angle of this discussion is discussed by O'Neil in her influential book *Weapons of Math Destruction* (O'Neil 2016).

## 2.4 Bias

Bias typically surfaces when unfair judgments are made because the individual making the judgment is influenced by a characteristic that is *actually* irrelevant to the matter at hand, typically a discriminatory preconception about members of a group. So, one form of bias is a learned cognitive feature of a person, often not made explicit. The person concerned may not be aware of having that bias – they may even be honestly and explicitly opposed to a bias they are found to have (e.g. through priming, cf. (Graham & Lowery 2004)).

Apart from the social phenomenon of learned bias, the human cognitive system is generally prone to have various kinds of 'cognitive biases', e.g. the 'confirmation bias': humans tend to interpret information as confirming what they already believe. This second form of bias is often said to impede performance in rational judgment (Kahnemann 2011) – though at least some cognitive biases generate an evolutionary advantage, e.g. economical use of resources for intuitive judgment. There is a question whether AI systems could or should have such cognitive bias.

A third form of bias is in present in data, when it exhibits systematic error, e.g. one of the various kinds of 'statistical bias'. Strictly, any given dataset will only be unbiased for a single kind of issue, so the mere creation of a dataset involves the danger that may it be used for a different kind of issue, and then turn out to be biased for that kind. Machine learning on the basis of such data would then not only fail to recognise the bias, but codify and automate the 'historical bias'. Such historical bias was discovered in an automated recruitment screening system at Amazon (discontinued early 2017) that discriminated against women – presumably because the company had a history of discriminating against women in the hiring process. The problem with such systems is thus their bias plus humans placing excessive trust in the systems. The political dimensions of such automated systems in the USA are investigated in

One use of prediction is in 'predictive policing', which many fear might lead to an erosion of public liberties because it can take away power from the people who's behaviour is predicted (Eubanks 2018; Ferguson 2017). It appears, however, that many of the worries about policing depend on futuristic scenarios where law enforcement foresees and punishes planned actions, rather than waiting until a crime has been committed (like in the 2002 film 'Minority Report'). Actual 'predictive policing' or 'intelligence led policing' techniques mainly concern the question of where and when police forces will be needed most – which is something a police force will always have done. Whether this is problematic depends on the appropriate level of trust in the technical quality of these systems, and on the evaluation of aims of the police work itself. Perhaps a recent paper title points in the right direction here: "AI ethics in predictive policing: From models of threat to an ethics of care" (Asaro 2019).

There are significant technical efforts to detect and remove bias from AI systems, but it is fair to say that these are in early stages: see UK Institute for Ethical AI & Machine Learning (Yeung & Lodge 2019). It appears that technological fixes have their limits in that they need a mathematical notion of fairness, which is hard to come by (Whittaker et al. 2018, 24ff).

### 2.5 Deception & Robots

Human-robot interaction (HRI) is an academic fields in its own right, which now pays significant attention to ethical matters, the dynamics of perception from both sides, and both the different interests present in and the intricacy of the social context, including co-working. Useful surveys for the ethics of robotics include (Calo, Froomkin, & Kerr 2016; Lin, Abney, & Jenkins 2017; Royakkers & van Est 2016).

While AI can be used to manipulate humans into believing and doing things, it can also be used to drive robots that are problematic if their processes or appearance involve deception, threaten human dignity, or violate the Kantian requirement of 're-spect for humanity'. Humans very easily attribute mental properties to objects, and empathise with them, especially when the outer appearance of these objects is similar to that of living beings. This can be used to deceive humans (or animals) into attributing more intellectual or even emotional significance to robots or AI systems than they deserve. Some parts of humanoid robotics are problematic in this regard (e.g. Hiroshi Ishiguro's remote-controlled Geminoids), and there are cases that have been clearly deceptive for public-relations purposes (e.g. Hanson Robotics' "Sophia"). Of course, some fairly basic constraints of business ethics and law apply to robots, too: product safety and liability, or non-deception in advertisement.

There are cases, however, where human-human interaction has aspects that appear specifically human in ways that can perhaps not be replaced by robots: care, love and sex.

The use of robots in health care for humans is currently at the level of concept studies in real environments, but it may become a usable technology in a few years, and has raised a number of concerns for a dystopian future of de-humanised care (Sparrow 2016). Current systems include robots that support human carers/caregivers (e.g. in lifting patients, or transporting material), robots that enable patients to do certain things by themselves (e.g. eat with a robotic arm), but also robots that are given to patients as company and comfort (e.g. the 'Paro' robot seal). For an overview, see (van Wynsberghe 2016). It is not very clear that there really is an issue here, since the discussion mostly focuses on the fear of robots de-humanising care, but the actual and foreseeable robots in care are for classic automation of technical tasks as assistive robots. They are thus 'care robots' only in a behavioural sense of performing tasks in care environments, not in the sense that a human 'cares' for the patients. Some robots that pretend to 'care' on a basic level are available (Paro seal) and others are in the making. A system that pretends to care would be deceptive and thus problematic – unless the deception is countered by sufficiently large utility gain. Perhaps feeling cared for by a machine can be progress in some cases?

Another area of discussion are sex robots: It has been argued by several tech optimists that humans will likely be interested in sex and companionship with robots and be comfortable with the idea (Levy 2007). Given the variation of human sexual preferences, including sex toys and sex dolls, this seems very likely: The question is whether such devices should be manufactured and promoted, and whether there should be limits to use (Danaher & McArthur 2017; Devlin 2018). Humans have long had deep emotional attachments to objects, so perhaps companionship or even love with a predictable android is attractive, especially to people who struggle with actual humans, and already prefer dogs, cats, a computer or a *tamagotchi*. In all this area there is an issue of deception, since a robot cannot (at present) mean what it says, or have feelings for a human. Having said that, paying for deception seems to be an elementary part of the traditional sex industry. Finally, there are concerns that have often accompanied matters of sex, namely consent, aesthetic concerns, and the worry that humans may be 'corrupted' by certain experiences. Old fashioned though this may seem, it is likely that pornography or sex robots support the perception of other humans as mere objects of desire, or even as recipients of abuse – and thus ruin a deeper sexual experience.

### 2.6 Autonomy & Responsibility

There are several notions of autonomy at play in the discussion of autonomous systems. A stronger notion is involved in philosophical debates where autonomy is the basis for responsibility and personhood. In this context, responsibility implies autonomy, but not inversely, so there can be systems that have degrees of technical autonomy without raising issues of responsibility. The weaker, more technical, notion of autonomy in robotics is relative and gradual: A system is said to be autonomous with respect to human control to a certain degree (Müller 2012).

Generally speaking, one question is the degree to which autonomous robots raise issues to which our present conceptual schemes must adapt, or whether they just require technical adjustments. In most jurisdictions, there is a sophisticated system of civil and criminal liability to resolve such issues. Technical standards, e.g. for the safe use of machinery in medical environments, will likely need to be adjusted. There is already a field of 'verifiable AI' for such safety-critical systems, and for 'security applications'. Technical Associations like the IEEE and the BSI have produced 'standards', particularly on more technical problems, such as data security and transparency. Among the many autonomous systems on land, on water, under water, in the air or in space, we discuss two samples: autonomous vehicles and autonomous weapons.

*Autonomous vehicles* hold the promise to reduce the very significant damage that human driving currently causes – with approximately 1 million humans being killed per year, many more injured, the environment polluted, soil sealed with tarmac, cities full of parked cars, etc. etc. However, there seem to be questions on how autonomous vehicles should behave, and how responsibility and risk should be distributed in the complicated system the vehicles operates in. There is some discussion of 'trolley problems' in this context. In the classic 'trolley problems' (Woollard & Howard-Snyder 2016, section 2) various dilemmas are presented: The simplest version is that of a trolley train on a track that is heading towards five people and will kill them, unless the train is diverted onto a side track, but on that track there is one person, who will be killed if the train takes that side track. 'Trolley problems' are not supposed to describe actual ethical problems or to be solved with a 'right' choice. Rather, they are thought-experiments where choice is artificially constrained to a small finite number of distinct one-off options and where the agent has perfect knowledge. These thought-experiments are used as a theoretical tool to investigate ethical intuitions and theories – especially the difference between actively doing vs. allowing something to happen, intended vs. tolerated consequences, and consequentialist vs. other normative approaches (Kamm & Rakowski 2016). This type of problem has reminded many of the problems encountered in actual driving, and in autonomous driving. It is doubtful, however, that an actual driver or autonomous car will ever have to solve trolley problems. While autonomous car trolley problems have received a lot of media attention, they do not seem to offer anything new to either ethical theory or to the programming of autonomous vehicles.

Our second example are *military robots*. The notion of automated weapons is fairly old: "For example, instead of fielding simple guided missiles or remotely piloted vehicles, we might launch completely autonomous land, sea, and air vehicles capable of complex, far-ranging reconnaissance and attack missions." (DARPA 1983, 1). This proposal was ridiculed as 'fantasy' at the time (Dreyfus, Dreyfus, & Athanasiou 1986, ix), but it is now a reality, at least for more easily identifiable targets (missiles, planes, ships, tanks, etc.). The main arguments against (lethal) autonomous weapon systems (AWS or LAWS), are that they support extrajudicial killings, take responsibility away from humans, and make wars or killings more likely (Lin, Bekey, & Abney 2008, 73-86). The crucial asymmetry where one side can kill with impunity, and thus has fewer reasons not to do so, already exists in conventional drone wars with remote controlled weapons (e.g. US in Pakistan). Another question seems to be whether using autonomous weapons in war would make wars worse, or perhaps make wars less bad? If robots reduce war crimes and crimes in war, the answer may well be positive and has been used as an argument in favour of these weapons (Müller 2016). Arguably the main threat is not the use of such weapons in conventional warfare, but in asymmetric conflicts or by non-state agents, including criminals. A lot has been made of keeping humans "in the loop" or "on the loop" in the military guidance on weapons (Santoni de Sio & van den Hoven 2018). There have been discussions about the difficulties of allocating responsibility for the killings of an autonomous weapon, and a 'responsibility gap' has been suggested (esp. Sparrow 2007), meaning that neither the human nor the machine may be responsible. On the other hand, we do not assume that for every event there is someone responsible for that event, and the real issue may well be the distribution of risk (Simpson & Müller 2016).

### 2.7   Singularity

In some quarters, the aim of current AI is thought to be an 'artificial general intelligence' (AGI) – this notion is usually distinguished from traditional notions of AI as a general purpose system, and from Searle's notion of 'strong AI': "computers given the right programs can be literally said to *understand* and have other cognitive states" (Searle 1980, 417).

The idea of the *singularity* is that if the trajectory of artificial intelligence towards AGI reaches up to systems that have a human level of intelligence, then these systems would themselves have the ability to develop AI systems that surpass the human level of intelligence, that is they are 'superintelligent' . Such superintelligent AI systems would quickly self-improve or develop even more intelligent systems (Chalmers 2010). This sharp turn of events after reaching superintelligent AI is the 'singularity', from where onwards the development of AI is out of human control and hard to predict (Kurzweil 2005, 487). (Bostrom 2014) explains in some detail what would happen at that point, and what the risks for humanity are.

The fear that "the robots we created will take over the world" had captured human imagination even before there were computers. It was first formulated by Irvin Good:

> Let an ultraintelligent machine be defined as a machine that can far surpass all the intellectual activities of any man however clever. Since the design of

> machines is one of these intellectual activities, an ultraintelligent machine could design even better machines; there would then unquestionably be an 'intelligence explosion', and the intelligence of man would be left far behind. Thus the first ultraintelligent machine is the last invention that man need ever make, provided that the machine is docile enough to tell us how to keep it under control. (Good 1965, 33).

The argument from acceleration to singularity is spelled out by Kurzweil (1999), who points out that computing power has been increasing exponentially, i.e. doubling ca. every 2 years since 1970 in accordance with 'Moore's Law' on the number of transistors, and will continue to do so for some time in the future. Kurzweil predicted that by 2010 supercomputers will reach human computation capacity, by 2030 'mind uploading' will be possible, and by 2045 the 'singularity' will occur. In addition to Moore's Law there is also an actual increase in the funds available to AI companies in recent years. There are possible paths to superintelligence other than computing power increase, e.g. the complete emulation of the human brain on a computer (Kurzweil 2012), biological paths, or networks and organisations (Bostrom 2014, 22-51). Despite obvious weaknesses in the identification of 'intelligence' with processing power, Kurzweil seems right that humans tend to underestimate the power of exponential growth.

The participants in this debate are united by being technophiles, in the sense that they expect technology to develop rapidly and bring broadly welcome changes – but beyond that, they divide into those that focus on benefits (e.g. Kurzweil) vs. those that focus on risks (e.g. Bostrom). Both camps sympathise with 'transhuman' views of survival for humankind in a different physical form, e.g. uploaded on a computer (Moravec 1990). They also consider the prospects of 'human enhancement', in various respects, including intelligence – often called "IA" (intelligence augmentation). It may be that future AI will be used for human enhancement, or will contribute further to the dissolution of the neatly defined human single person.

The argument from superintelligence to risk requires the assumption that superintelligence does not imply benevolence – contrary to Kantian traditions in ethics that have argued higher levels of rationality or intelligence would go along with a better understanding of what is moral, and better ability to act morally (Chalmers 2010, 36f). Arguments for risk from superintelligence say that rationality and morality are entirely independent dimensions – this is sometimes explicitly argued for as an "orthogonality thesis" (Bostrom 2014, 105-109).

Criticism of the singularity narrative has been raised from various angles. Kurzweil and Bostrom seem to assume that intelligence is a one-dimensional property and that the set of intelligent agents is well-ordered in the mathematical sense – but neither discusses intelligence at any length in their books. Generally, it is fair to say that despite some efforts, the assumptions made in the powerful narrative of superintelligence and singularity have not been investigated in detail. Philosophically, one interesting question is whether singularity is on the trajectory of actual AI research, or not (e.g. Brooks 2017; Müller forthcoming). This discussion raises the question whether the concern about 'singularity' is just a narrative about fictional AI based on human fears. But even if one *does* find negative reasons compelling and the singularity not likely to occur, there is still a significant possibility that one may turn out to be wrong. So, it appears that discussion of the very high-impact risk of singularity has justification *even if* one thinks the probability of such singularity ever occurring is very low.

Thinking about superintelligence in the long term raises the question whether superintelligence may lead to the extinction of the human species, which is called an "existential risk" (or XRisk): The superintelligent systems may well have preferences that conflict with the existence of humans on Earth, and may thus decide to end that existence – and given their superior intelligence, they will have the power to do so (or they may happen to end it because they do not really care). These issues are sometimes taken more narrowly to be about human extinction (Bostrom 2013), or more broadly as concerning any large risk for the species – of which AI is only one (Häggström 2016).

In a narrow sense, the 'control problem' is how we humans can remain in control of an AI system once it is superintelligent (Bostrom 2014, 127ff). In a wider sense it is the problem how we can make sure an AI system will turn out to be positive, in the sense we humans perceive this (Russell 2019); this is sometimes called 'value alignment'. How easy or hard it is to control a superintelligence depends to a significant extent on the speed of 'take-off' to a superintelligent system. One aspect of this problem is that we might decide a certain feature is desirable, but then find out that it has unforeseen consequences that are so negative that we would not desire that feature after all. This is the ancient problem of King Midas who wished that all he touched would turn into gold.

Discussions about superintelligence include speculation about omniscient beings, the radical changes on a 'latter day', and the promise of immortality through transcendence of our current bodily form – so they have clear religious undertones (Geraci 2010). These issues also pose a well-known problem of epistemology: Can we know the ways of the omniscient? Opponents would thus say we need an ethics for the 'small' problems that occur with actual AI & robotics, and less for the 'big ethics' of existential risk from AI.

### 2.8    *Machine Ethics*

Machine ethics is ethics for machines, for 'ethical machines', for machines as *subjects*, rather than for the human use of machines as *objects*. It is often not very clear whether this is supposed to cover all of AI ethics or to be a part of it (Floridi & Saunders 2004; Moor 2006; Wallach & Asaro 2017). Sometimes it looks as though there is the dubious inference at play here that if machines act in ethically relevant ways, then we need a machine ethics. Some of the discussion in machine ethics makes the very substantial assumption that machines can, in some sense, be ethical agents responsible for their actions, or 'autonomous moral agents'. It is not clear that there is a consistent notion of 'machine ethics' since weaker versions are in danger of reducing 'having an ethics' to notions that would not normally be considered sufficient (e.g. without 'reflection' or even without 'action'); stronger notions that move to-wards artificial moral agents may describe a – currently – empty set. So, in this article, we discuss ethical issues (above) and the notion of moral agency in artificial systems (below), but we do not expand a separate discussion of machine ethics.

### 2.9 Artificial Moral Agents

If one takes machine ethics to concern moral agents, in some substantial sense, then these agents can be called 'artificial moral agents', having rights and responsibilities. However, the discussion about artificial entities challenges a number of common notions in ethics and it can be very useful to understand these in abstraction from the human case (cf. Powers & Ganascia forthcoming). If the robots act, will they themselves be responsible, liable or accountable for their actions? Or should the distribution of risk perhaps take precedence over discussions of responsibility?

Several authors use 'artificial moral agent' in a less demanding sense, borrowing from the use of 'agent' in software engineering, in which case matters of responsibility and rights will not arise. James Moor (2006) distinguishes four types of machine agents: ethical impact agents (example: robot jockeys), implicit ethical agents (example: safe autopilot), explicit ethical agents (example: using formal methods to estimate utility), and full ethical agents ("can make explicit ethical judgments and generally is competent to reasonably justify them. An average adult human is a full ethical agent.") Several ways to achieve 'explicit' or 'full' ethical agents have been proposed, via programming it in (operational morality), via 'developing' the ethics itself (functional morality) and finally full-blown morality with full intelligence and sentience (Allen, Smit, & Wallach 2005; Moor 2006).

In some of these discussions the notion of 'moral patient' plays a role: Ethical *agents* have responsibilities while ethical *patients* have rights, because harm to them matters. It seems clear that some entities are patients without being agents, e.g. simple animals that can feel pain but cannot make justified choices. On the other hand it is normally understood that all agents will also be patients (e.g. in a Kantian framework). Usually, being a person is supposed to be what makes an entity a responsible agent, someone who can have duties and be the object of ethical concerns, and such personhood is typically a deep notion associated with free will (Strawson 2004) and with having phenomenal consciousness.

Traditional distribution of responsibility already occurs: A car maker is responsible for the technical safety of the car, a driver is responsible for driving, a mechanic is responsible for proper maintenance, the public authorities are responsible for the technical conditions of the roads, etc. In general "The effects of decisions or actions based on AI are often the result of countless interactions among many actors, including designers, developers, users, software, and hardware. … With distributed agency comes distributed responsibility." (Taddeo & Floridi 2018, 751). How this distribution might occur is not a problem that is specific to AI.

Some authors have indicated that it should be seriously considered whether current robots must be allocated rights (Danaher 2019; Gunkel 2018). This position seems to rely largely on criticism of the opponents and on the empirical observation that robots and other non-persons are sometimes treated as having rights. In this vein, a 'relational turn' has been proposed: If we relate to robots as though they had rights, then we might be well-advised not to search whether they 'really' do have such rights (Coeckelbergh 2010). This raises the question how far such anti-realism or quasi-realism can go, and what it means then to say that 'robots have rights' in a human-centred approach.

There is a wholly separate issue whether robots (or other AI systems) should be given the status of 'legal entities', or 'legal persons' – in a sense in which natural persons, but also states, businesses or organisations are 'entities', namely they can have legal rights and duties. The European Parliament has considered allocating such status to robots in order to deal with civil liability (Parliament 2016), but not criminal liability – which is reserved for natural persons. It would also be possible to assign only a certain subset of rights and duties to robots. In environmental ethics there is a long-standing discussion about the legal rights for natural objects like trees (Stone 1972).

In the community of 'artificial consciousness' researchers there is a significant concern whether it would be ethical to create such consciousness, since creating it would presumably imply ethical obligations to a sentient being, e.g. not to harm it and not to end its existence by switching it off – some authors have called for a "moratorium on synthetic phenomenology" (Bentley, Brundage, Häggström, & Metzinger 2018, 28f).

## 3     Future Developments

The singularity thus raises the problem of the concept of AI again. It is remarkable how imagination or 'vision' has played a central role since the very beginning of the discipline at the 'Dartmouth Summer Research Project' (1956). And the evaluation of this vision is subject to dramatic change: In a few decades, we went from the slogans "AI is impossible" (Dreyfus 1972) and "AI is just automation" (Lighthill 1973) to "AI will solve all problems" (Kurzweil 1999) and "AI may kill us all" (Bostrom 2014). This created media attention and PR efforts, but it also raises the problem how much of this 'philosophy and ethics of AI' is really about AI, rather than about an imagined technology. – As we said at the outset, AI and robotics have raised fundamental questions about what we should do with these systems, what the systems themselves should do, and what risks they have in the long term. They also challenge the human view of humanity as the intelligent and dominant species on Earth. We have seen issues that have been raised and we will have to watch technological and social developments closely to catch the new issues early on, and to develop a philosophical analysis, as well as to learn for traditional problems of philosophy.

## 4     References

Abowd, J. M. (2017). How will statistical agencies operate when all data are private? *Journal of Privacy and Confidentiality, 7* (3), 1-15.

AI4EU. (2019). Outcomes from the Strategic Orientation Workshop (Deliverable 7.1). (June 28, 2019). ai4eu.eu

AI HLEG. (2019). High-level expert group on artificial intelligence: Ethics guidelines for trustworthy AI. *European Commission, 09.04.2019.* https://ec.europa.eu/digital-single-market/en/high-level-expert-group-artificial-intelligence

Allen, C., Smit, I., & Wallach, W. (2005). Artificial Morality: Top-down, Bottom-up, and Hybrid Approaches. *Ethics and Information Technology, 7* (3), 149-155. doi: 10.1007/s10676-006-0004-4

Asaro, P. M. (2019). AI ethics in predictive policing: From models of threat to an ethics of care. *IEEE Technology and Society Magazine, 38* (2), 40-53. doi: 10.1109/MTS.2019.2915154

Bennett, C. J., & Raab, C. (2006). *The governance of privacy: Policy instruments in global perspective* (2nd ed.). Cambridge, Mass.: MIT Press.

Bentley, P. J., Brundage, M., Häggström, O., & Metzinger, T. (2018). Should we fear artificial intelligence? In-depth analysis. *European Parliamentary Research Service, Scientific Foresight Unit (STOA), March 2018*(PE 614.547), 1-40. http://www.europarl.europa.eu/RegData/etudes/IDAN/2018/614547/EPRS_IDA%282018%29614547_EN.pdf

Bostrom, N. (2013). Existential risk prevention as global priority. *Global Policy, 4* (1), 15-31.

Bostrom, N. (2014). *Superintelligence: Paths, dangers, strategies*. Oxford: Oxford University Press.

Brooks, R. (2017, 07.09.2017). The seven deadly sins of predicting the future of AI. from https://rodneybrooks.com/the-seven-deadly-sins-of-predicting-the-future-of-ai/

Brundage, M., Avin, S., Clark, J., Toner, H., Eckersley, P., Garfinkel, B., . . . Filar, B. (2018). The malicious use of artificial intelligence: Forecasting, prevention, and mitigation. *FHI/CSER/CNAS/EFF/OpenAI Report*, 1-101. https://arxiv.org/abs/1802.07228

Bryson, J. J. (2019). The past decade and future of AI's impact on society. In Anonymous (Ed.), *Towards a new enlightenment: A transcendent decade*. Madrid: Turner - BVVA.

Calo, R. (2018). Artificial Intelligence Policy: A Primer and Roadmap *University of Bologna Law Review, 3* (2), 180-218. doi: http://dx.doi.org/10.2139/ssrn.3015350

Calo, R., Froomkin, M. A., & Kerr, I. (Eds.). (2016). *Robot law*). Cheltenham: Edward Elgar.

Chalmers, D. J. (2010). The singularity: A philosophical analysis. *Journal of Consciousness Studies, 17* (9-10), 7-65.

Coeckelbergh, M. (2010). Robot rights? Towards a social-relational justification of moral consideration. *Ethics and Information Technology, 12* (3), 209-221. doi: 10.1007/s10676-010-9235-5

Costa, E., & Halpern, D. (2019). The behavioural science of online harm and manipulation, and what to do about it: An exploratory paper to spark ideas and debate. *The Behavioural Insights Team Report*, 1-82. https://www.bi.team/publications/the-behavioural-science-of-online-harm-and-manipulation-and-what-to-do-about-it/

Danaher, J. (2019). Welcoming Robots into the Moral Circle: A Defence of Ethical Behaviourism. *Science and Engineering Ethics*. doi: 10.1007/s11948-019-00119-x

Danaher, J., & McArthur, N. (Eds.). (2017). *Robot sex: Social and ethical implications*). Boston, Mass.: MIT Press.

DARPA. (1983). Strategic computing - new-generation computing technology: A strategic plan for its development an application to critical problems in defense (28.10.1983). from http://www.scribd.com/document/192183614/Strategic-Computing-1983

Devlin, K. (2018). *Turned on: Science, sex and robots*. London: Bloomsbury.

Dreyfus, H. L. (1972). *What computers still can't do: A critique of artificial reason* (2 ed.). Cambridge, Mass.: MIT Press 1992.

Dreyfus, H. L., Dreyfus, S. E., & Athanasiou, T. (1986). *Mind over machine: The power of human intuition and expertise in the era of the computer*. New York: Free Press.

Eubanks, V. (2018). *Automating inequality: How high-tech tools profile, police, and punish the poor*. London: St. Martin's Press.

European Group on Ethics in Science and New Technologies. (2018, 09.03.2018). Statement on artificial intelligence, robotics and 'autonomous' systems. *European Commission, Directorate-General for Research and Innovation, Unit RTD.01.* from http://ec.europa.eu/research/ege/pdf/ege_ai_statement_2018.pdf

Ferguson, A. G. (2017). *The rise of big data policing: Surveillance, race, and the future of law enforcement.* New York: NYU Press.

Floridi, L., Cowls, J., Beltrametti, M., Chatila, R., Chazerand, P., Dignum, V., . . . Vayena, E. (2018). AI4People—An ethical framework for a good AI society: Opportunities, risks, principles, and recommendations. *Minds and Machines, 28* (4), 689-707.

Floridi, L., & Saunders, J. W. (2004). On the Morality of Artificial Agents'. *Minds and Machines, 14*, 349-379.

Floridi, L., & Taddeo, M. (2016). What is Data Ethics? *Phil. Trans. R. Soc. A, 374* (2083).

GDPR. (2016). General Data Protection Regulation: Regulation (EU) 2016/679 of the European Parliament and of the Council of 27 April 2016 on the protection of natural persons with regard to the processing of personal data and on the free movement of such data, and repealing Directive 95/46/EC. *Official Journal of the European Union, 119*(04.05.2016), 1–88. http://data.europa.eu/eli/reg/2016/679/oj

Geraci, R. M. (2010). *Apocalyptic AI: Vision of heaven in robotics, artificial intelligence and virtual reality.* Oxford: Oxford University Press.

Gibert, M. (2019). Éthique artificielle (version Grand Public). In M. Kristanek (Ed.), *Encyclopédie Philosophique.*

Good, I. J. (1965). Speculations concerning the first ultraintelligent machine. In F. L. Alt & M. Ruminoff (Eds.), *Advances in Computers* (Vol. 6, pp. 31-88). New York & London: Academic Press.

Graham, S., & Lowery, B. S. (2004). Priming Unconscious Racial Stereotypes About Adolescent Offenders. *Law and Human Behavior, 28* (5), 483-504. doi: 10.1023/B:LAHU.0000046430.65485.1f

Gunkel, D. J. (2018). The other question: can and should robots have rights? *Ethics and Information Technology, 20* (2), 87–99. doi: 10.1007/s10676-017-9442-4

Gunning, D. (2018). Explainable artificial intelligence (XAI). *Defense Advanced Research Projects Agency.* from https://www.darpa.mil/program/explainable-artificial-intelligence

Häggström, O. (2016). *Here be dragons: Science, technology and the future of humanity.* Oxford: Oxford University Press.

Harari, Y. N. (2016). *Homo deus: A brief history of tomorrow.* New York: Harper.

Harris, T. (2016). How technology is hijacking your mind—from a magician and google design ethicist. *medium.com, Thrive Global* (18.05.2016).

Jobin, A., Ienca, M., & Vayena, E. (2019). The global landscape of AI ethics guidelines. *Nature Machine Intelligence, 1* (9), 389-399. doi: 10.1038/s42256-019-0088-2

Kahnemann, D. (2011). *Thinking fast and slow.* London: Macmillan.

Kamm, F. M., & Rakowski, E. (Eds.). (2016). *The trolley problem mysteries*). New York: Oxford University Press.

Kissinger, H. A. (2018). How the enlightenment ends: Philosophically, intellectually—in every way—human society is unprepared for the rise of artificial intelligence. *The Atlantic, June.* https://www.theatlantic.com/magazine/archive/2018/06/henry-kissinger-ai-could-mean-the-end-of-human-history/559124/

Kurzweil, R. (1999). *The age of spiritual machines: When computers exceed human intelligence.* London: Penguin.

Kurzweil, R. (2005). *The singularity is near: When humans transcend biology.* London: Viking.

Kurzweil, R. (2012). *How to create a mind: The secret of human thought revealed.* New York: Viking.

Levy, D. (2007). *Love and sex with robots: The evolution of human-robot relationships*. New York: Harper & Co.

Lighthill, J. (1973). Artificial intelligence: A general survey. *Artificial intelligence: A paper symposion*, (London). http://www.chilton-computing.org.uk/inf/literature/reports/lighthill_report/p001.htm

Lin, P., Abney, K., & Jenkins, R. (Eds.). (2017). *Robot ethics 2.0: From autonomous cars to artificial intelligence*). New York: Oxford University Press.

Lin, P., Bekey, G., & Abney, K. (2008). Autonomous military robotics: Risk, ethics, and design. *US Department of Navy, Office of Naval Research* (December 20, 2008), 1-112.

Macnish, K. (2017). *The ethics of surveillance: An introduction*. London: Routledge.

Mathur, A., Acar, G., Friedman, M., Lucherini, E., Mayer, J., Chetty, M., & Narayanan, A. (2019). Dark patterns at scale: Findings from a crawl of 11K shopping websites. *Proceedings of the ACM Human-Computer Interaction, 3* (81), 1-32.

Moor, J. H. (2006). The nature, importance, and difficulty of machine ethics. *IEEE Intelligent Systems, 21* (4), 18-21.

Moravec, H. (1990). *Mind children*. Cambridge, Mass.: Harvard University Press.

Müller, V. C. (2012). Autonomous cognitive systems in real-world environments: Less control, more flexibility and better interaction. *Cognitive Computation, 4* (3), 212-215. doi: 10.1007/s12559-012-9129-4

Müller, V. C. (2016). Autonomous killer robots are probably good news. In E. Di Nucci & F. Santoni de Sio (Eds.), *Drones and responsibility: Legal, philosophical and socio-technical perspectives on the use of remotely controlled weapons* (pp. 67-81). London: Ashgate.

Müller, V. C. (forthcoming). *Can machines think? Fundamental problems of artificial intelligence*. New York: Oxford University Press.

O'Neil, C. (2016). *Weapons of math destruction: How big data increases inequality and threatens democracy*. Largo, ML: Crown.

Parliament, E. (2016, 31.05.2016). Draft report with recommendations to the Commission on Civil Law Rules on Robotics (2015/2103(INL)). *Committee on Legal Affairs*.

Powers, T. M., & Ganascia, J.-G. (forthcoming). The ethics of the ethics of AI. In M. D. Dubber, F. Pasquale & S. Das (Eds.), *Oxford Handbook of Ethics of Artificial Intelligence*. New York.

Roessler, B. (2017). Privacy as a human right. *Proceedings of the Aristotelian Society, 2* (CXVII).

Royakkers, L., & van Est, R. (2016). *Just ordinary robots: Automation from love to war*. Boca Raton: CRC Press, Taylor & Francis.

Russell, S. (2019). *Human compatible: Artificial intelligence and the problem of control*. New York: Viking.

Santoni de Sio, F., & van den Hoven, J. (2018). Meaningful Human Control over Autonomous Systems: A Philosophical Account. *Frontiers in Robotics and AI, 5* (15). doi: 10.3389/frobt.2018.00015

Schneier, B. (2015). *Data and Goliath: The hidden battles to collect your data and control your world*. New York: W. W. Norton.

Searle, J. R. (1980). Minds, brains and programs. *Behavioral and Brain Sciences, 3*, 417-457.

Sennett, R. (2018). *Building and dwelling: Ethics for the city*. London: Allen Lane.

Simpson, T. W., & Müller, V. C. (2016). Just war and robots' killings. *The Philosophical Quarterly, 66* (263), 302-322. doi: 10.1093/pq/pqv075

Sparrow, R. (2007). Killer robots. *Journal of Applied Philosophy, 24* (1), 62-77.

Sparrow, R. (2016). Robots in aged care: A dystopian future. *AI & SOCIETY, 31* (4), 1-10.

Stone, C. D. (1972). Should trees have standing - Toward legal rights for natural objects. *Southern California Law Review* (2), 450-501.

Strawson, G. (2004, 29.02.2004). Free will. *Routledge Encyclopedia of Philosophy*. Retrieved May 2005, 2005, from http://www.rep.routledge.com/article/V014

Susser, D., Roessler, B., & Nissenbaum, H. (2019). Technology, autonomy, and manipulation. *Internet Policy Review, 8* (2). doi: 10.14763/2019.2.1410

Taddeo, M., & Floridi, L. (2018). How AI can be a force for good. *Science, 361* (6404), 751-752. doi: 10.1126/science.aat5991

van Wynsberghe, A. (2016). *Healthcare robots: Ethics, design and implementation*. London: Routledge.

Wachter, S., Mittelstadt, B. D., & Russell, C. (2018). Counterfactual explanations without opening the black box: Automated decisions and the GDPR. *Harvard Journal of Law & Technology, 31* (2).

Wallach, W., & Asaro, P. M. (Eds.). (2017). *Machine Ethics and Robot Ethics*). London: Routledge.

Whittaker, M., Crawford, K., Dobbe, R., Fried, G., Kaziunas, E., Mathur, V., . . . Schultz, J. (2018). AI Now Report 2018. from https://ainowinstitute.org/AI_Now_2018_Report.html

Woollard, F., & Howard-Snyder, F. (2016). Doing vs. allowing harm. *Stanford Encyclopedia of Philosophy*.

Woolley, S., & Howard, P. (Eds.). (2017). *Computational propaganda: Political parties, politicians, and political manipulation on social media*). Oxford: Oxford University Press.

Yeung, K., & Lodge, M. (Eds.). (2019). *Algorithmic regulation*). Oxford: Oxford University Press.

Zerilli, J., Knott, A., Maclaurin, J., & Gavaghan, C. (2019). Transparency in algorithmic and human decision-making: Is there a double standard? *Philosophy & Technology, 32* (4), 661–683. doi: 10.1007/s13347-018-0330-6

Zuboff, S. (2019). *The age of surveillance capitalism: The fight for a human future at the new frontier of power*. New York: Public Affairs.